\documentclass{article}
\usepackage{spconf,amsmath,graphicx}
\usepackage{booktabs}
\usepackage{makecell}
\usepackage{multirow}
\usepackage{hyperref}

\title{DEEP CNN BASED FEATURE EXTRACTOR FOR TEXT-PROMPTED SPEAKER RECOGNITION}

\name{Sergey Novoselov$^{1,2}$, Oleg Kudashev$^2$, Vadim Schemelinin$^1$, Ivan Kremnev$^3$, Galina Lavrentyeva$^1$}

\address{
  $^1$ITMO University, St.Petersburg, Russia \\
  $^2$STC-innovations Ltd., St.Petersburg, Russia\\
  $^3$STC Ltd., St.Petersburg, Russia\\
\\
  \texttt{\{novoselov,kudashev,shchemelinin,kremnev,lavrentyeva\}@speechpro.com}}

\begin{document}
%
\maketitle
\begin{abstract}
Deep learning is still not a very common tool in speaker verification field.
We study deep convolutional neural network performance
in the text-prompted speaker verification task.
The prompted passphrase is segmented into word states --- i.e. digits --- to test each digit utterance separately. We train a single high-level feature extractor for all states and use cosine similarity metric for scoring.
The key feature of our network is the Max-Feature-Map activation function, which acts as an embedded feature selector. By using multitask learning scheme to train the high-level feature extractor we were able to surpass the classic baseline systems in terms of quality and achieved impressive results for such a novice approach, getting 2.85\% EER on the \textit{RSR2015} evaluation set. Fusion of the proposed and the baseline systems improves this result.
\end{abstract}
\begin{keywords}
speaker verification, text-prompted, deep features, CNN, max-feature mapping  
\end{keywords}

\section{Introduction}
\label{sec:intro}

I-vector-based systems are well known to be state-of-the-art solutions to the text-independent speaker verification problem \cite{kenny-fa, novel, novoselov_non_linear_plda}.
Nonetheless, this problem is gradually gaining attention from the deep learning perspective. Particularly, studies \cite{novel, novoselov_dnn} make use of the ASR deep neural network (ASR DNN) in order to divide acoustic space into senone classes, and the classic total variability (TV) model is applied to discriminate between speakers in that space afterwards \cite{kenny-fa}. 

In such phonetic discriminative DNN-based systems two major techniques can be distinguished. The first one uses DNN posteriors to calculate Baum-Welch statistics, and the second one uses bottleneck features in pair with speaker specific features (MFCC) for a full TV-UBM system training.

Recent publications \cite{larcher, hagay, tdjfa, novoselov_state_plda} suggest that substantial advancement of the state-of-the-art text-dependent verification systems is mainly based on progress in the text-independent speaker recognition task. Therefore, successful application of a phonetic discriminative DNN to the latter task justifies employment of a similar approach for text-dependent systems \cite{matejka, zeinali_hmm_ivec, td-prelim}. 

On the other hand, direct speaker discrimination is the most natural way of speaker verification.
There are several profound studies on advantageous usage of deep end-to-end solutions for discriminating speakers directly in a text-dependent task \cite{microsoft, google}. 
Paper \cite{dnnsmall} describes a DNN that extracts a small speaker footprint that is used to discriminate between speakers. Paper \cite{snyder2017deep} presents a well performing implementation of a DNN extractor based on the speaker discriminative approach in the text-independent task.

This paper focuses on the text-prompted speaker verification scenario. It does not require user to remember a specific passphrase and reduces risk of replay attacks on the system. Previously, we have shown that by segmenting a passphrase into word states prior to supervector extraction we were able to construct more accurate statistical models of speech signals \cite{tdjfa, novoselov_state_plda}. The present study continues with this approach and suggest ways of further improvement of deep speaker verification systems.

This work exploits the deep convolutional neural network (CNN) with Max-Feature-Map activation function (MFM) from \cite{lightcnn, lavrentyeva2017audio}, which is based on maxout activation function, for direct speaker discrimination in the text-dependent setting.
We train a deep high-level feature extractor for prompted passwords. 
The experiments were mainly conducted on Part 3 of the \textit{RSR2015} database \cite{rsr2015}, which contains series of randomized digit sequences representing passwords. The results of the proposed and the baseline systems are compared in section~\ref{sec:results}.

In addition, we analyze performance of the proposed system on the training data extended with English (\textit{Wells Fargo Bank}) and Russian (\textit{STC-Russian-digits}) digits subcorpora.

\section{Baseline systems}
\label{sec:baseline}

This section briefly describes three state-of-the-art text-dependent speaker verification systems, which are further referred as baseline systems. To do passphrase segmentation, we use hidden Markov model (HMM) based Viterbi alignment. The frontend of the baseline systems computes mel-frequency cepstral coefficients (MFCC) as well as their first and second time derivatives to yield a 39-dimensional vector per frame. Framing is done every 8 ms with a 16 ms window. A gaussian mixture model (GMM) based voice activity detector is
used to find and remove non-speech frames. We also apply
cepstral mean subtraction (CMS) and do not apply feature
warping to cepstral coefficients.

\subsection{GMM-SVM}
\label{sec:gmm_svm}

The baseline GMM-SVM system is implemented in accordance to \cite{hagay}. The  GMM mean supervector $\vec{m}$ is obtained from a relevant maximum a posteriori (MAP) adaptation of the speaker-independent universal background model (UBM) of a passphrase. No segmentation is done by this system.
We trained a passphrase UBM on the \textit{RSR2015} database development set. To compensate for inter-speaker variability, Nuisance Attribute Projection (NAP) is applied. Support Vector Machine (SVM) is used as a backend classifier. Finally, score s-normalization is done.

\subsection{State-GMM-SVM}
\label{sec:stgmm_supervec}

In this system HMM segmentation is used to split a passphrase into individual digits.
For each digit a State-GMM mean supervector is extracted as described in \cite{tdjfa}. Each state is associated with a unique speaker-independent UBM, which is trained on the \textit{RSR2015} database training set. The speaker-dependent GMM means of a state are obtained through MAP adaptation of the UBM means. The speaker-dependent GMM mean supervector $\vec{m}$ is obtained with concatenation of the speaker state mean vectors over all states. NAP, SVM and score s-normalization are also utilized in this system.

We refer to this system as the StGMM-SVM system.

\subsection{STATE-PLDA}
\label{sec:plda}

We have previously shown effectiveness of the State probabilistic linear discriminant analysis (State-PLDA) model when addressing the text-dependent speaker verification problem \cite{novoselov_state_plda}. 
%
%

Log-likelihood ratio score for each digit in a tested utterance can be obtained with PLDA scoring procedure. The final score for the utterance is calculated as a sum of the \textit{llr} scores over its states \cite{novoselov_state_plda}.
No score normalization is needed.

This system is called StPLDA.

\section{State CNN}
\label{sec:cnn}

Encouraged by the success of our deep learning model in the replay attack spoofing detection task \cite{lavrentyeva2017audio}, we endeavour to improve over the state-of-the-art techniques for text-prompted speaker verification by employing a similar network architecture.

\subsection{Features}
\label{sec:features}
Our CNN-based system also operates with distinct states, and the same HMM-based segmentation is used to split passphrase into separate digits.
We use log mel power spectra extracted from the speech signal as input features. While the feature size along the frequency domain axis is fixed to $64$ bands, varying utterance time span must be adjusted to a fixed length of $96$ frames, which is an estimation of the longest digit pronouncement time. This is done either by cropping the end of features along the time axis or wrap padding as it was done in \cite{lavrentyeva2017audio}. Cepstral mean and variance normalization is done for each digit.

\subsection{MFM CNN architecture}
\label{sec:architecture}

We make use of the Light CNN architecture \cite{lightcnn}. MFM activation function is applied to the feature maps inferred from a convolutional layer:
\begin{align}
\begin{split}
    y_{i,j}^k = \max (x_{i,j}^k,\, x_{i,j}^{k + \frac{N}{2}}) \\
    i = \overline{1,W}, \; j = \overline{1, H}, \; k = \overline{1,\frac{N}{2}}
\end{split}
\end{align}
with $x$ being an MFM layer input tensor of shape $W \times H \times N$ and $y$ its output tensor of shape $W \times H \times \frac{N}{2}$. Here $i$, $j$ indicates the frequency and time domains and $k$ is the channel index.

The network itself is composed of several layer groups, each of them is represented by a convolutional layer followed by an MFM layer and optionally a pooling layer. These groups are stacked together and followed by dense layers to produce embeddings for input log mel power spectra features. Details are covered in Table~\ref{tab:cnn}.

The choice of the MFM activation function is motivated by the fact that it not only provides ReLU-like activation with a varying threshold but is also an embedded feature selector \cite{lightcnn}.

\subsection{Training}
\label{sec:training}

There are two ways of training a deep neural network extractor in the text-prompted scenario. The first one is to train the network to discriminate between speakers (single task mode) and the second, more complex one, is to train the network to discriminate between speakers and digits simultaneously (multitask mode). The latter approach is used in \cite{dnnmultitask}.
We also exploit the multitask setup and assign an individual class to a speaker pronouncing a particular digit. Because of this total amount of classes increases by a factor of 10, i.e. $N_{speakers} \times N_{digits}$.
The network is trained in multitask mode with multiclass cross entropy loss function.

\subsection{Scoring}
\label{sec:scoring}
After training, the last fully-connected layer with its softmax activation is removed from the network in order to obtain an extractor of high-level representations for speaker pronounced digits. 
At the evaluation step embeddings for each individual digit utterance are obtained through the network forward pass. Cosine similarity metric is used afterwards to compare them with corresponding targets.
Each test digit embedding is compared to the same digit embedding from the enrollment set, which is averaged over 3 enrollment sessions.
Finally, average score of all digits in the passphrase is returned. Note that we do not use any discriminant analysis backend instead of cosine similarity.

\section{Experimental setup}
\label{sec:expsetup}

\subsection{Databases description}
\label{sec:database}

We have conducted experiments on Part 3 of the \textit{RSR2015} database \cite{rsr2015} and on the evaluation part of the \textit{STC-Russian-digits} database (STCRus). These datasets contain prompted passphrases for speaker verification.

The training set is also extended with \textit{Wells Fargo Bank} dataset (\textit{WF}), described in \cite{tdjfa}, which contains short digit passphrase utterances, and the training part of \textit{STC-Russian-digits} dataset.

\textit{STC-Russian-digits} is a newly collected database. It contains utterances of random Russian digit sequences pronounced by native speakers. Each speaker has recorded 1 to 3 sessions, each of them consists of microphone records made with 3 different devices: an Android-based mobile phone, an iOS-based mobile phone, and a web-camera. Each record includes 5 pronunciations of 10 Russian digits in random order. \textit{STC-Russian-digits} database is split into training and evaluation parts according to Table~\ref{tab:stcdigits}.
\begin{table}[b]
  \caption{STC-Russian-digits database}
  \label{tab:stcdigits}
  \centering
  \begin{tabular}{|c|c|c|c|c|}
    \hline
       Set &  Males & Females & Total speakers & Total records \\
    \hline
       Train & 473 & 221 & 694 & 15 k \\ 
       Eval & 50 & 42 & 92 & 5 k \\
    \hline
  \end{tabular}
\end{table}

\subsection{CNN parameters}
\label{sec:params}

Network parameters are described in Table~\ref{tab:cnn}. The last dense layer is included only at the training stage. The number of its neurons corresponds to the number of classes in a particular setup, described in~\ref{sec:training}.

\begin{table}[t]
  \caption{CNN architecture}
  \label{tab:cnn}
  \centering
  \begin{tabular}{l l l r}
    \toprule
    \textbf{Type} & \textbf{Filter / Stride} & \textbf{Output} & \textbf{\#Params}   \\
    \midrule
    Conv1      & $7 \times 7$ / $1 \times 1$ & $64 \times 96 \times 128$ & 6.4K      \\
    MFM1    & $-$                         & $64 \times 96 \times 64$ & $-$      \\
    \midrule
    MaxPool1   & $2 \times 2$ / $2 \times 2$ & $32 \times 48 \times 64$ & $-$      \\
    \midrule
    Conv2a     & $1 \times 1$ / $1 \times 1$ & $32 \times 48 \times 128$ & 8.3K      \\
    MFM2a   & $-$                         & $32 \times 48 \times 64$ & $-$      \\
    Conv2b     & $5 \times 5$ / $1 \times 1$ & $32 \times 48 \times 192$ & 153.8K     \\
    MFM2b   & $-$                         & $32 \times 48 \times 96$ & $-$      \\
    \midrule
    MaxPool2   & $2 \times 2$ / $2 \times 2$ & $16 \times 24 \times 96$ & $-$      \\
    \midrule
    Conv3a     & $1 \times 1$ / $1 \times 1$ & $16 \times 24 \times 192$ & 18.6K     \\
    MFM3a   & $-$                         & $16 \times 24 \times 96$ & $-$      \\
    Conv3b     & $5 \times 5$ / $1 \times 1$ & $16 \times 24 \times 256$ & 614.7K    \\
    MFM3b   & $-$                         & $16 \times 24 \times 128$ & $-$      \\
    \midrule
    MaxPool3   & $2 \times 2$ / $2 \times 2$ & $8 \times 12  \times 128$ & $-$      \\
    \midrule
    Conv4a     & $1 \times 1$ / $1 \times 1$ & $8 \times 12  \times 256$ & 33K     \\
    MFM4a   & $-$                         & $8 \times 12  \times 128$ & $-$      \\
    Conv4b     & $3 \times 3$ / $1 \times 1$ & $8 \times 12  \times 128$ & 147.6K     \\
    MFM4b   & $-$                         & $8 \times 12  \times 64$ & $-$      \\
    \midrule
    Conv5a     & $1 \times 1$ / $1 \times 1$ & $8 \times 12  \times 128$  & 8.3K      \\
    MFM5a   & $-$                         & $8 \times 12  \times 64$  & $-$      \\
    Conv5b     & $3 \times 3$ / $1 \times 1$ & $8 \times 12  \times 128$  & 73.9K     \\
    MFM5b   & $-$                         & $8 \times 12  \times 64$  & $-$      \\
    \midrule
    MaxPool4   & $2 \times 2$ / $2 \times 2$ & $4 \times 6  \times 64$  & $-$      \\
    \midrule
    FC1        & $-$                         & $2048$              & 3147.8K     \\
    MFM6    & $-$                         & $1024$                       & $-$      \\
    \midrule
    FC2         & $-$                        & $N_{out}$              & $1024 N_{out}$      \\
    \midrule
    Total       &                         &               & \thead{4366K +\\ $1024N_{out}$}     \\
    \bottomrule
\end{tabular}
\end{table}

The learning rate is decayed by a constant every 10 epochs. Batch size is set to 32.

\section{Results and discussion}
\label{sec:results}

Tables \ref{tab:comparison}, \ref{tab:lang} and \ref{tab:fusion} present equal error rate (EER) and minimum detection cost function (minDCF) \footnote{$P_{\textrm{tar}} = 10^{-2}$, $C_{\textrm{miss}} = 10$ and $C_{\textrm{fa}} = 1$} for various systems. The best performance among baseline systems is demonstrated by \textit{StGMM-SVM}, yielding 3.11\% EER for pooled male and female conditions. The speaker discriminative deep CNN trained in the single task mode (\textit{$StCNN^{ST}$}) could not surpass baseline results, achieving only 7.83\% EER. However, it showed significant performance improvement in multitask mode (\textit{$StCNN^{MT}$}) with EER decreased to 5.12\%. 
In contrast to \textit{GMM-SVM} and \textit{StGMM-SVM}, no score normalization is needed for \textit{StPLDA} and \textit{StCNN} based systems.

\begin{figure}[h]
  \centering
  \includegraphics[width=0.9\linewidth]{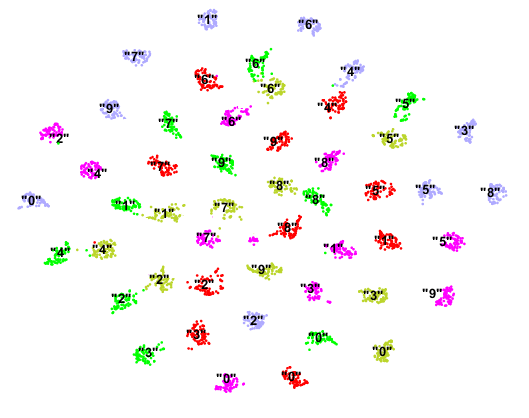}
  \caption{Visualizing speakers embeddings using t-SNE. Each speaker is marked by its color, text indicates pronounced digit}
  \label{fig:tsne}
\end{figure}

Remarkably, when new language corpora with foreign classes are presented in the training set, the system learns to discriminate embeddings between speakers for both languages, based on common speaker variability (see Table \ref{tab:lang}). Two bottom lines in Table \ref{tab:comparison} show that the extending of the training set for the CNN-based system leads to performance boost. The best performing system \textit{$StCNN^{MT}$} \big[+WF+STCRus\big] we have obtained had been trained on the pooled English (\textit{RSR2015} and \textit{WF}) and Russian (\textit{STC-Russian-digits}) subcorpora.
Unlike the baseline systems, which benefit from training on in-domain data only, the deep CNN-based system improves from new out-of-domain data added during the learning phase in the multitask setup. 
Figure \ref{fig:tsne} illustrates embedding discriminative capability of the system: 5 randomly chosen speaker embeddings are projected on two principal axes with t-SNE \cite{tsne}.

\begin{table}[]
\centering
\caption{EER [\%] (minDCF [\%]) for 5-digit password verification. Here \textit{tr} and \textit{ev} indices stand for training and evaluation parts of the used data bases.}
\label{tab:comparison}
\begin{tabular}{| c | c | c | c |}
\hline
System & Male & Female & Pooled \\
\hline
\thead{\textit{$GMM$-$SVM$}} & \small 4.01 (17,91) & \small 2.58 (14.69) & \small 3.37 (16.45) \\
\hline
\thead{\textit{$StGMM$-$SVM$}} & \small 3.65 (16.29) & \small 2.13 (1.1.06) & \small 3.11 (14.23) \\
\hline
\thead{\textit{$StPLDA$}} & \small 3.37 (17.42) & \small 3.04 (16.1) & \small 4.44 (21.94) \\
\hline
\thead{\textit{$StCNN^{ST}$}} & \small 7.06 (37.13) & \small 8.4 (40.62) & \small 7.83 (39.1) \\
\hline
\thead{\textit{$StCNN^{MT}$}} & \small 4.88 (23.6) & \small 5.19 (24.82) & \small 5.12 (24.51) \\
\hline
\thead{\textit{$StCNN^{MT}$}\\ \big[+$WF$\big]} & \small 4.16 (19.53) & \small 4.23 (19.9) & \small 4.27 (20.03) \\
\hline
\thead{\textit{$StCNN^{MT}$} \\ \big[+$WF$+$STCRus_{tr}$\big]} & \small 2.68 (12.31) & \small 3.13 (14.55) & \small 2.85 (13.36) \\
\hline
\end{tabular}
\end{table}

\setlength{\tabcolsep}{4pt}
\begin{table}[]
\centering
\caption{EER [\%] (minDCF [\%]) for pooled male-female case (5-digit password)}
\label{tab:lang}
\begin{tabular}{| c | c | c | c | }
\hline
System    & TrainData                     & TestData     & EER (minDCF) \\
\hline       
\multirow{6}{*}{\small\textit{$StCNN^{MT}$}}  &\small $RSR2015_{tr}$ & \multirow{3}{*}{\small $RSR2015_{ev}$} & \small 5,12 (24,51)\\
                                        \cline{2-2} \cline{4-4}
                                        & \thead{$RSR2015_{tr}$  + \\$WF$ + $STCRus_{tr}$} & & 2,85 (13,36) \\
                                        \cline{2-4}
                                        & \small $STCRus_{tr}$ & \multirow{3}{*}{\small $STCRus_{ev}$} & 5,86 (29,49)\\
                                        \cline{2-2} \cline{4-4} 
                                        & \thead{$RSR2015_{tr}$  + \\$WF$ + $STCRu_{tr}$} &              & 4,24 (20,45) \\
\hline
\end{tabular}
\end{table}
There are two advantages of using speaker and text discriminative embeddings, trained on English and Russian digits speech signals. First, it automatically validate correctness of the passphrase. Second, it can be used for both languages (see Table \ref{tab:lang}). 

Table \ref{tab:fusion} shows evaluation metrics for the best \textit{StCNN} system fused with baseline systems. Fusion of all systems enabled us to reduce verification EER on the pooled male and female 5-digit passphrase test to 1.43\%, which is 54\% less than the best single baseline system. Fusion is done with BOSARIS toolkit \cite{brummer}.

\begin{table}[h]
\centering
\caption{Fusion. EER [\%] (minDCF [\%]) for 5-digit password verification}
\label{tab:fusion}
\begin{tabular}{| c | c | c | c |}
\hline
System                        & Male        & Female        & Pooled        \\
\hline
\thead{\textit{StCNN} +\\ \textit{StPLDA}} & 2.05 (9.12)   & 1.77 (8.62)   & 2.09 (9.59)   \\
\hline
\thead{\textit{StCNN} +\\\textit{StGMM-SVM}}   & 1.61 (7.7)    & 1.4  (6.43)   & 1.63 (7.32)   \\
\hline
\thead{\textit{StCNN} +\\ \textit{GMM-SVM}}     & 1.48 (6.62) & 1.56 (8.66) & 1.57 (7.72)  \\
\hline
All                           & 1.26 (5.02)   & 1.31 (6.92)   & 1.43 (6.58)  \\
\hline
\end{tabular}
\end{table}

\section{Conclusion}
\label{sec:conclusion}

    This paper proposed a new deep CNN-based solution to the text-prompted speaker verification problem. A high level feature extractor was trained to discriminate between speakers and digits simultaneously, and embeddings can be measured simply with cosine similarity. By using multitask learning scheme we were able to surpass the classic baseline systems in terms of quality and achieved impressive results for such a novice approach, obtaining 2.85\% EER on the \textit{RSR2015} evaluation set. Fusion of the proposed and the baseline systems led to almost 50\% decrease in EER for all test sets.

\textbf{Acknowledgements}. This work was financially supported by the Ministry of Education and Science of the Russian Federation, Contract 14.578.21.0189 \\(ID RFMEFI57816X0189).

\vfill\pagebreak


\bibliographystyle{IEEE}

\end{document}